\documentclass[transmag]{IEEEtran}
\usepackage{latexsym}
\usepackage{graphicx}
\usepackage{amsfonts,amssymb,amsmath}
\usepackage{hyperref}

\usepackage{xcolor}
\usepackage{amsmath}
\usepackage{CJK}
\usepackage{url}
\usepackage{array}

\usepackage{ragged2e}
\usepackage[misc]{ifsym}

\ifCLASSOPTIONcompsoc
  \usepackage[nocompress]{cite}
\else
  \usepackage{cite}
\fi

\ifCLASSINFOpdf
\else
\fi

\ifCLASSOPTIONcompsoc
  \usepackage[caption=false,font=footnotesize,labelfont=sf,textfont=sf]{subfig}
\else
  \usepackage[caption=false,font=footnotesize]{subfig}
\fi

\def\BibTeX{{\rm B\kern-.05em{\sc i\kern-.025em b}\kern-.08em T\kern-.1667em\lower.7ex\hbox{E}\kern-.125emX}}
\begin{document}

\title{Network Resource Allocation Strategy Based on Deep Reinforcement Learning}

\author{Shidong Zhang, Chao Wang, Junsan Zhang, Youxiang Duan, Xinhong You, and Peiying Zhang

\thanks{This work is partially supported by the Major Scientific and Technological Projects of CNPC under Grant ZD2019-183-006, and partially supported by ``the Fundamental Research Funds for the Central Universities" of China University of Petroleum (East China) under Grant 20CX05017A, 18CX02139A. \textit{Corresponding authors: Peiying Zhang and Chao Wang.}}
\thanks{Shidong Zhang is with State Grid Shandong Electric Power Research Institute, Jinan, 250003, China. (e-mail: huanazhang@163.com)}
\thanks{Chao Wang, Junsan Zhang and Youxiang Duan are with the College of Computer Science and Technology, China University of Petroleum (East China), Qingdao 266580, China. (e-mail: 774248254@qq.com, zhangjunsan@upc.edu.cn, yxduan@upc.edu.cn)}
\thanks{Xinhong You is with State Grid Shandong Electric Power Research Institute, Jinan, 250003, China. (e-mail: youxh93@163.com)}
\thanks{Peiying Zhang is with the College of Computer Science and Technology, China University of Petroleum (East China), Qingdao 266580, China. (e-mail: zhangpeiying@upc.edu.cn).}}

\IEEEtitleabstractindextext{\begin{abstract}
The traditional Internet has encountered a bottleneck in allocating network resources for emerging technology needs. Network virtualization (NV) technology as a future network architecture, the virtual network embedding (VNE) algorithm it supports shows great potential in solving resource allocation problems. Combined with the efficient machine learning (ML) algorithm, a neural network model close to the substrate network environment is constructed to train the reinforcement learning agent. This paper proposes a two-stage VNE algorithm based on deep reinforcement learning (DRL) (TS-DRL-VNE) for the problem that the mapping result of existing heuristic algorithm is easy to converge to the local optimal solution. For the problem that the existing VNE algorithm based on ML often ignores the importance of substrate network representation and training mode, a DRL VNE algorithm based on full attribute matrix (FAM-DRL-VNE) is proposed. In view of the problem that the existing VNE algorithm often ignores the underlying resource changes between virtual network requests, a DRL VNE algorithm based on matrix perturbation theory (MPT-DRL-VNE) is proposed. Experimental results show that the above algorithm is superior to other algorithms.
\end{abstract}

}

\maketitle

\section{INTRODUCTION}

\IEEEPARstart{T}{he} rapid development of the Internet in the past few decades has made great contributions to the progress of human society \cite{Z1}. However, with the rise of new technologies such as cloud computing, the Internet of Things (IoT) and 5G, the traditional Internet architecture has been unable to meet its explosive development \cite{B1,B2}. In particular, the contradiction between the expanding functional requirements of network users and the inefficient and unreasonable allocation of underlying network resources in Internet infrastructure is more significant \cite{B3}. It is mainly reflected in the following aspects: First, the development of emerging network paradigms such as the IoT and 5G requires strong and robust basic network architecture as support. They need the underlying network to provide a large amount of network resources in a short time to support its operation without causing network congestion or even paralysis. Secondly, with the explosive growth of network end users, users' network requirements should also be differentiated. For example, in the field of Internet of Vehicles (IoV), driverless vehicles require extremely low-latency networks to ensure that they can accurately determine road conditions in real time \cite{D1,a2}. In the business of information backup, a lot of bandwidth resources are needed to ensure the integrity and security of a large amount of data backup. In view of the above differentiated quality of service (QoS) requirements, the existing Internet architecture cannot reasonably allocate the underlying resources to them, which is very easy to make the final allocation result fall into the local optimal situation \cite{Y1}. Finally, the Internet architecture only provides a "best effort" service delivery model. The allocation of the underlying network resources is random, and it is easy for resources to be fragmented. There are problems such as unreasonable resource allocation and low resource utilization \cite{Z2,D2}. Radio network resource management faces severe challenges, including storage, spectrum, computing resource allocation, and joint allocation of multiple resources \cite{jcx1,jcx2}. With the rapid development of communication networks, the integrated space-ground network has also become a key research object \cite{jcx3}.

The emergence of NV provides a reliable solution to the problem of resource allocation, which is considered to be a future network architecture with broad application prospects \cite{B4,Z3}. In a NV environment, the resource point-to-point transmission bandwidth is large, the delay is low and the stability is high. Based on these advantages, the NV architecture can serve the emerging work of the IoT, artificial intelligence and 5G communications \cite{D3,D4}. VNE as the key to NV should be focused on. In recent years, using DRL algorithm to improve the efficiency of VNE has become a research hotspot. Therefore, this paper focuses on the application of DRL to VNE algorithm.

NV refers to the abstraction of the underlying physical resources into multiple virtual networks to realize the sharing of the underlying physical resources. How to map nodes and links in virtual network to nodes and links in substrate network under constraint conditions (resource constraint, location constraint, etc.) is the VNE problem \cite{a1,Z4,B5}. Most of the existing VNE algorithms are designed based on heuristic methods, which have the following limitations: VNE tends to converge to the local optimal solution and cannot obtain the global optimal solution; The heuristic approach relies on manually formulating a set of rules and assumptions that do not fully reproduce the reality of substrate network and virtual network and the connections between them \cite{B6,B7}.

DRL algorithm is a new ML algorithm, which is composed of deep learning algorithm and reinforcement learning algorithm. The DRL algorithm mainly solves the decision problem in high dimensional space and state space \cite{D5,B8,Y2}. In deep learning algorithm, it mainly relies on artificially constructed neural networks to solve high-dimensional decision-making problems. It is an algorithmic mathematical model that imitates the behavioral characteristics of animal neural networks and performs parallel processing of distributed information. The more classic artificial neural networks include perceptron, BP neural network, radial basis neural network and feedback neural network \cite{B9,B10}. The main idea of reinforcement learning is to use an agent to learn (train) the decision-making strategy in the process of interacting with the environment, so as to obtain the optimal reward \cite{1,a3,B11}. Since the VNE problem is proved to be NP-hard, using DRL method to solve the VNE problem is an effective way.

In recent years, due to the rapid development of cloud computing services, the world's major Internet service providers in order to obtain greater profits, constantly update and improve the network architecture and service quality. Therefore, the application of DRL to solve the VNE problem has become a hot topic \cite{B12,B13}. Many researchers at home and abroad have also proposed many ML-based VNE algorithms. However, due to the initial stage of this technology, various theoretical ideas need to be improved urgently. Therefore, the proposed algorithm generally has some disadvantages, such as low mapping efficiency, large consumption of resources and not meeting the actual network situation. Therefore, aiming at these problems, this paper discusses several more perfect VNE algorithms based on DRL \cite{Y3}. The results show that it can effectively make up for the shortcomings of existing algorithms.

\begin{figure*}[!htbp]
\centering
\includegraphics[width=0.90\textwidth]{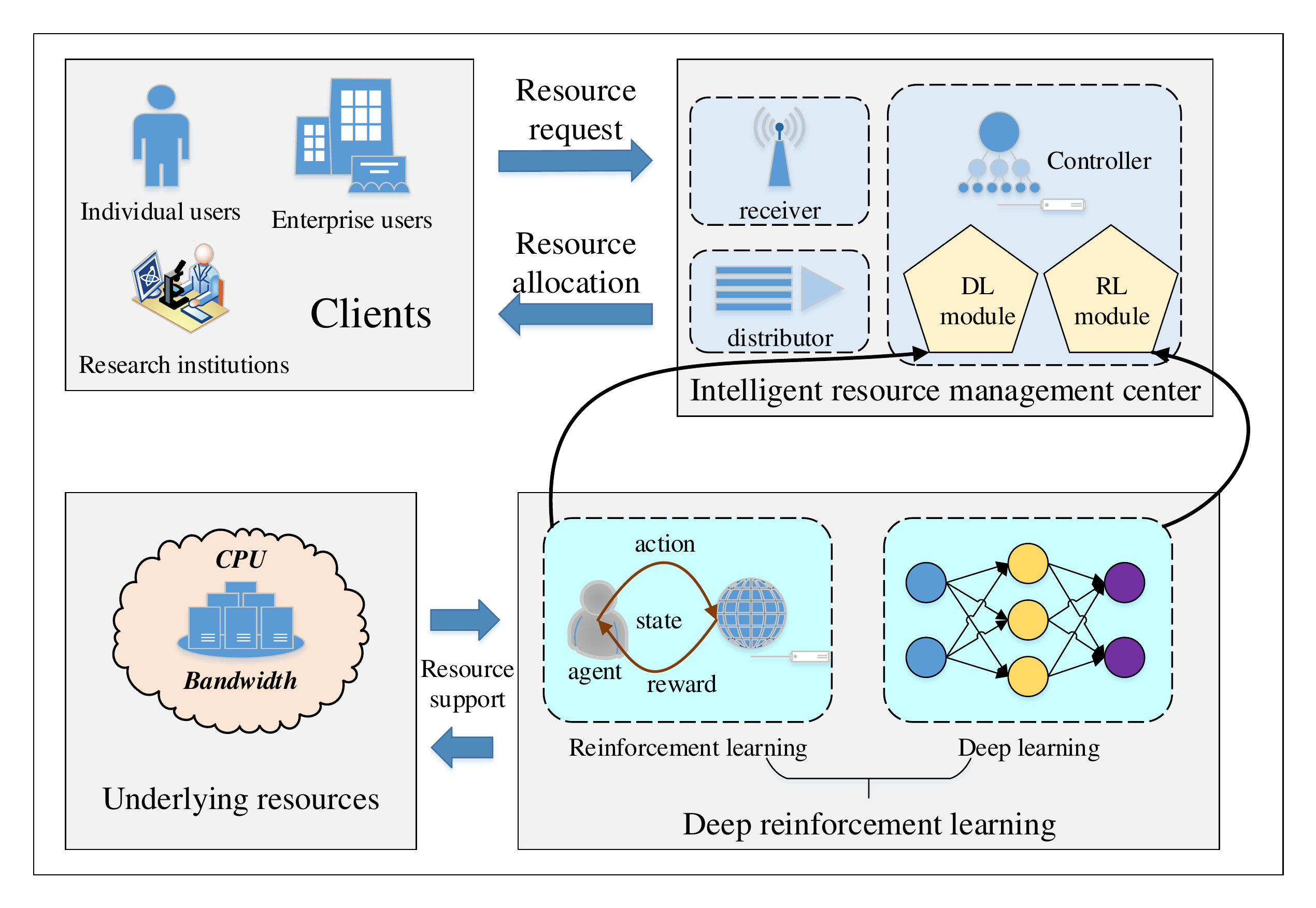}
\caption{Resource allocation framework based on deep reinforcement learning.}
\label{fig_1}
\end{figure*}

The resource allocation framework based on DRL is shown in Figure \ref{fig_1}. Network end user design has a wide range, as long as the network users who need to request network resources can be considered as end users. It mainly includes individual users, enterprise users and research institutions. The network end user sends the virtual network resource request to the network service center. The resource manager is located in the intelligent management center of network resources, receives the network resource requests from the end users through the receiving port and allocates the network resources to different end users through the distribution port. The scheduling of network resources is completed by the combination of deep learning and reinforcement learning. DRL agent is trained in different network environment and the optimal resource allocation strategy is deduced according to different objectives. The underlying resources are distributed in the underlying resource pool, mainly including CPU resources and bandwidth resources.

The structure of this paper is organized as follows: The first part is an overview of the problems of VNE and DRL, which leads to the problem of combining the two. The second part describes the basic problems of VNE, then establishes the relevant network model. The third part puts forward the constraints of VNE, then establishes the evaluation indexes used in the experimental part. In the forth part, the TS-DRL-VNE algorithm is proposed for the case that the mapping results of heuristic algorithms tend to converge to the local optimal solution. In the fifth part, the FAM-DRL-VNE algorithm is proposed to solve the problem of ignoring the importance of substrate network representation and training mode. In the sixth part, the MPT-DRL-VNE algorithm is proposed, aiming at the problem that the DRL-based VNE algorithm often ignores the change of the underlying resources in the stage of virtual network request cohesion. The last part summarizes the whole paper.

Specifically, We list in TABLE \uppercase\expandafter{\romannumeral1} the main ideas of the three algorithms proposed in this article and the several algorithms to be compared later.

\begin{table*}
\centering
\caption{Algorithm ideas}
\renewcommand\arraystretch{1.5}
\begin{tabular}{|p{30mm}|p{130mm}|}
\hline
The algorithm name & The algorithm description\\
\hline
TS-DRL-VNE & The VNE problem is transformed into a combinatorial optimization problem. The node mapping strategy is established by using a pointer network containing two long and short time memory modules. Then, an active search algorithm based on policy gradient is designed to optimize the pointer network without preprocessing virtual network requests and searching for the optimal strategy. \\
\hline
FAM-DRL-VNE & The method adopts substrate network spectrum analysis. The attribute matrix and the adjacency matrix are considered together to obtain a robust full attribute matrix that can represent the substrate network. Then the VNE model was trained by DRL. The agent can effectively discover the relationship between substrate network and virtual network requests, thus completing the VNE process. \\
\hline
MPT-DRL-VNE & The algorithm uses the matrix perturbation theory to capture the change of substrate network at the continuous time node. An efficient updating method of substrate network feature representation is completed. Then the VNE model was trained by DRL. The agent can effectively discover the relationship between substrate network and virtual network requests, thus completing the VNE process.\\
\hline
VNE-PTIC \cite{5} & The virtual topology pre-transformation mechanism is used to reduce topology differences and achieve fairness of virtual network request acceptance. The VNE problem is modeled as an integer linear programming problem and an algorithm based on discrete particle swarm optimization is used to solve the problem. \\
\hline
NodeRank \cite{B14} & A Markov random walk model is used to define the importance of a network node and rank it. The virtual nodes are mapped to the substrate nodes according to the ranking of the virtual nodes. Solve the problem of multi-commodity flow with divisible paths by searching for the shortest path with indivisible paths and complete the mapping of virtual links. \\
\hline
BaseLine \cite{B15} & Use the formula $H(n^s)=CPU(n^s)\sum_{l^s \in L(n^s)}BW(l^s)$ to sort the bottom nodes and map the bottom nodes with high ranking first. The shortest path algorithm is used in the link mapping phase. \\
\hline
RLVNE \cite{9} & Using a policy network as a reinforcement learning agent, the node mapping strategy is developed by training the agent. The policy network is trained by using historical data based on virtual network requests and strategy gradients are used to automatically optimize. Then use the breadth-first strategy to complete the virtual link mapping. \\
\hline
\end{tabular}
\end{table*}

\section{Problem Description and Network Model}

\subsection{Virtual Network Embedding Problem Description}

\begin{figure}[!htbp]
\includegraphics[width=1.0\columnwidth]{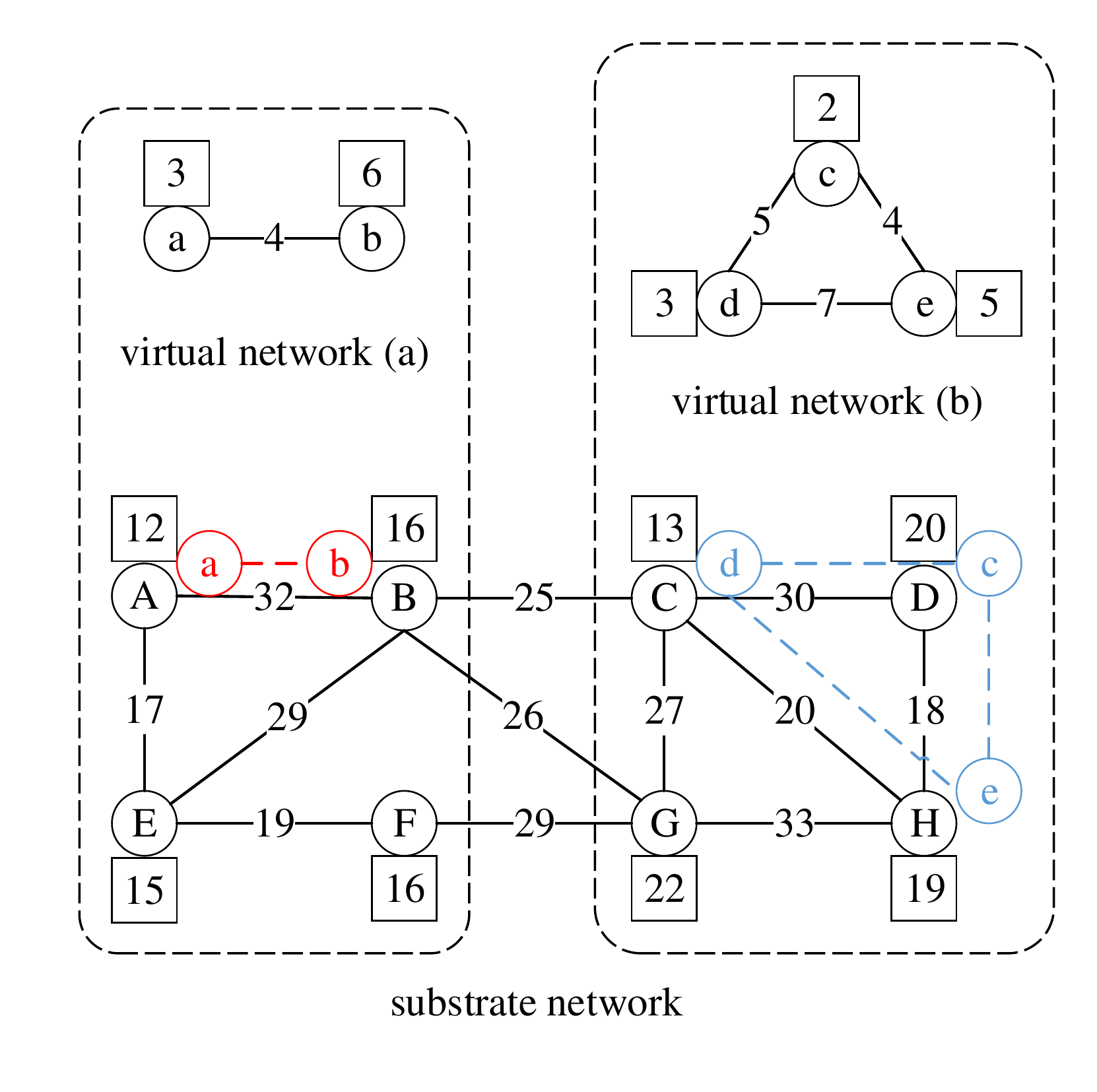}
\caption{Virtual networks embedded in the substrate network.}
\label{fig_2}
\end{figure}

Different network function requirements of users can be regarded as multiple heterogeneous virtual networks. The rational allocation of the underlying network resources for these virtual networks is called VNE \cite{Z5}. The underlying network resources usually include CPU resources of the underlying network nodes and bandwidth resources of the underlying network links. Figure \ref{fig_2} shows the mapping of two heterogeneous virtual networks to the underlying network.

The virtual network \textit{(a)} and virtual network \textit{(b)} are composed of two and three virtual nodes, respectively. The links between virtual nodes represent virtual links. The number in the square represents the CPU resource demand of the virtual node. The number on the virtual link represents the bandwidth resource requirement of the link. The substrate network is also composed of substrate nodes and substrate links. The number in the square represents the amount of CPU resources that the substrate node can provide. The number on the substrate link represents the amount of bandwidth resources that the link can provide. Figure \ref{fig_2} shows the successful mapping of two virtual networks, where the node correspondence are $a \to A, b \to B, c \to D, d \to C, e \to H$.

\subsection{Network Model}

\subsubsection{Substrate Network Model}

The substrate network is modeled as a weighted undirected graph represented by $G^S=\{N^S,L^S,CPU(N^S),BW(L^S)\}$. $N^S$ represents all substrate nodes. $L^S$ is a collection of all substrate links. $CPU(N^S)$ represents the CPU resource attribute of the substrate node. $BW(L^S)$ represents the bandwidth resource attribute of the substrate link. A substrate node in the substrate network is usually denoted by $n^s$ and $n^s \in N^S$. A substrate link in the substrate network is usually denoted by $l^s$ and $l^s \in L^S$.

\subsubsection{Virtual Network Model}

The virtual network is also modeled as a weighted undirected graph represented by $G^V=\{N^V,L^V,CPU(N^V),BW(L^V)\}$. $N^V$ represents all virtual nodes. $L^V$ is a collection of all virtual links. $CPU(N^V)$ represents the demand attribute of CPU resources of virtual nodes. $BW(L^V)$ represents the demand attribute of the bandwidth resource of virtual links. A virtual node in the virtual network is usually denoted by $n^v$ and $n^v \in N^V$. A virtual link in a virtual network is usually denoted by $l^v$ and $l^v \in L^V$.

After the substrate network model and the virtual network model are established, the process of VNE can be expressed as $G^V \to G^S$. One virtual node in the same virtual network can only be mapped to one substrate node. A substrate node can only carry one virtual node in the same virtual network request, that is, the two are in a one-to-one correspondence. A virtual link can be mapped to one or more substrate links, and the latter has a link split situation.

We summarize the network symbols used in TABLE \uppercase\expandafter{\romannumeral2}.

\begin{table}
\caption{Network symbols}
\renewcommand\arraystretch{1.5}
\begin{tabular}{|p{17mm}|p{63mm}|}
\hline
$G^S$ & entire substrate network \\
\hline
$N^S$ & substrate node set \\
$L^S$ & substrate link set \\
$n^s$ & a substrate node \\
$l^s$ & a substrate link \\
$CPU(n^s)$ & CPU resource of substrate node $n^s$ \\
$BW(l^s)$ & bandwidth resource of substrate link $l^s$ \\
\hline
$G^V$ & entire virtual network \\
\hline
$N^V$ & virtual node set \\
$L^V$ & virtual link set \\
$n^v$ & a virtual node \\
$l^v$ & a virtual link \\
$CPU(n^v)$ & CPU resource demand of virtual node $n^v$ \\
$BW(l^v)$ & bandwidth resource demand of virtual link $l^v$ \\
\hline
$G^V \to G^S$ & virtual network $G^V$ is mapped to substrate network $G^S$ \\
\hline
$n^v \to n^s$ & virtual node $n^v$ is mapped to substrate node $n^s$ \\
$l^v \to l^s$ & virtual link $l^v$ is mapped to substrate link $l^s$ \\
\hline
\end{tabular}
\end{table}

\section{Constraints and Evaluation Indicators}

\subsection{Constraints on the Embedding Problem of Virtual Networks}

Because the resources of the underlying network are limited, virtual network requests cannot be mapped onto the underlying network without restriction. How to use the limited underlying resources and allocate these resources to as many virtual network requests as possible is one of the most important goals of the VNE problem.

In general, the following constraints need to be considered during the VNE process.

\begin{equation}
\begin{aligned}
CPU(n^s) \ge CPU(n^v),
\end{aligned}
\end{equation}
means that the virtual node $n^v$ is mapped to the substrate node $n^s$, and the available CPU resource amount of the substrate node $n^s$ is not less than the CPU resource requirement of the virtual node $n^v$.

\begin{equation}
\begin{aligned}
BW(l^s) \ge BW(l^v),
\end{aligned}
\end{equation}
means that the virtual link $l^v$ is mapped onto the substrate link $l^s$, and the available bandwidth resource of the substrate link $l^s$ is not less than the bandwidth resource requirement of the virtual link $l^v$.

\begin{equation}
\begin{aligned}
\forall \, n^s \in N^S,n^v \in N^V \, num(n^v \to n^s)=1,
\end{aligned}
\end{equation}
means that for the virtual node $n^v$ in the same virtual network request, it can only be mapped to one substrate node $n^s$. The two are in one-to-one correspondence.

\begin{equation}
\begin{aligned}
\forall \, l^s \in L^S,l^v \in L^V \, num(l^v \to l^s) \ge 1,
\end{aligned}
\end{equation}
means that a virtual link $l^v$ can be mapped to one substrate link $l^s$ or multiple substrate links. The latter is a case of virtual link segmentation.

\subsection{Evaluation Index of Virtual Network Embedding Problem}

In the VNE problem, the currently available CPU resources of substrate node $n^s$ is usually expressed by the remaining CPU resources:

\begin{equation}
\begin{aligned}
R_{CPU}^{n^s} = CPU(n^s) - \sum_{all\,n^v \to n^s}CPU(n^s),
\end{aligned}
\end{equation}
where $R_{CPU}^{n^s}$ represents the current remaining CPU resource amount of the substrate node $n^s$. $CPU(n^s)$ represents the initial CPU resource amount of the substrate node $n^s$. $\sum_{all\,n^v \to n^s}CPU(n^s)$ represents the sum of CPU resources consumed by all virtual nodes $n^v$ mapped to substrate nodes $n^s$. The symbol $n^v \to n^s$ indicates that the virtual node $n^v$ is mapped onto the substrate node $n^s$.

The amount of currently available bandwidth resources of a physical link is usually expressed in terms of remaining bandwidth resources:

\begin{equation}
\begin{aligned}
R_{BW}^{l^s} = BW(l^s) - \sum_{all\,l^v \to l^s}BW(l^s),
\end{aligned}
\end{equation}
where $R_{BW}^{l^s}$ represents the current remaining bandwidth resource amount of the substrate link $l^s$. $BW(l^s)$ represents the initial bandwidth resource amount of the substrate link $l^s$. $\sum_{all\,l^v \to l^s}BW(l^s)$ represents the sum of bandwidth resources consumed by mapping all virtual links $l^v$ onto substrate links $l^s$. The symbol $l^v \to l^s$ indicates that the virtual link $l^v$ is mapped onto the substrate link $l^s$.

In the VNE problem, several indexes such as long-term average revenue, long-term revenue consumption ratio, link utilization rate and virtual network request acceptance rate are usually used to evaluate the performance of the VNE algorithm. In this paper, in order to reflect the advantages of the algorithm, we mainly use the index of long-term revenue consumption ratio to evaluate this paper and other algorithms. To some extent, the long-term revenue consumption ratio reflects the utilization degree of the underlying resources. If the revenue consumption is high, the utilization rate of the underlying resources is high and the number of virtual network requests accepted is also large. Therefore, the index has certain representativeness.

Before defining the long-term revenue consumption ratio, it needs to first define the long-term revenue embedded in the virtual network:

\begin{equation}
\begin{aligned}
R(G^V,t)=\sum_{n^v \in N^V}CPU(n^v)+\sum_{l^v \in L^V}BW(l^v).
\end{aligned}
\end{equation}

It can be seen from the above formula that the benefits of embedding virtual network $G^V$ into the substrate network are determined by the sum of the CPU resources consumed by the virtual node and the bandwidth resources consumed by the virtual link. Where $t$ is the duration of the virtual network $G^V$ request.

The consumption of VNE is:

\begin{equation}
\begin{aligned}
C(G^V,t)=\sum_{n^v \in N^V}CPU(n^v)+\sum_{l^s \in L^S}\sum_{l^v \in L^V}BW(f_{l^s}{l^v}),
\end{aligned}
\end{equation}
where $BW(f_{l^s}{l^v})$ represents the total bandwidth consumed by the virtual link $l^v$ mapped to the substrate link $l^s$. In the process of link mapping, virtual link $l^v$ may have link segmentation, which will be allocated to multiple substrate links, so the total consumed bandwidth needs to be calculated.

Therefore, the long-term income consumption ratio of the embedded virtual network is defined as:

\begin{equation}
\begin{aligned}
R/C=\lim_{T \to \infty}\frac{\sum_{t=0}^{T}R(G^V,t)}{\sum_{t=0}^{T}C(G^V,t)}.
\end{aligned}
\end{equation}

\section{Two Stages Virtual Network Embedding Algorithm Based on DRL}

At present, the solution of VNE based on DRL is still in its infancy. Most studies that use ML algorithms to solve VNE problems only use one of deep learning or RL alone. The use of heuristic methods to solve VNE problems is still the mainstream. However, one of the disadvantages of using heuristic methods to solve VNE problems is that the mapping results tend to converge to the local optimal solution and fail to reach the global optimal. Therefore, it has been unable to meet the requirements of modern network architecture. Therefore, we modeled the VNE problem as a combinatorial optimization problem and proposed a DRL-based coordination two-stage VNE algorithm.

The existing heuristic based VNE algorithms can be divided into two types: single-stage embedding and two-stage embedding. In reference \cite{2}, a hybrid integer structuring algorithm was proposed. The algorithm deals with node mapping and link mapping cooperatively, that is, single-stage embedding. The process of node mapping and link mapping is modeled by hybrid integer programming and solved in polynomial time. Reference \cite{3} uses the Vhub linear programming algorithm. The VNE problem is mapped to a mixed integer programming problem using the p-hub median method. The virtual node is modeled as a hub, and as long as the hub location is resolved, it can serve as the best mapping location for each virtual request. Finally, this method realizes the load balancing of single-stage VNE. In reference \cite{4}, path separation and migration algorithm was proposed. This algorithm rethinks the ability of substrate network so that substrate network can be used more efficiently. Then the path segmentation and path migration underlying link mapping strategy is proposed, which effectively utilizes the physical bandwidth and improves the robustness of the mapping strategy.

However, the above heuristic algorithms generally have the disadvantage that they cannot converge to the optimal solution. In order to improve the shortcomings of heuristic based VNE algorithm, we modeled the VNE problem as a combinatorial optimization problem and proposed a TS-DRL-VNE algorithm. In the two stages of virtual node mapping and virtual link mapping, DRL is applied to train agents in a large amount under the training set. The purpose is to select the physical node and physical link with the highest mapping probability. The main idea is: As more and more virtual network requests arrive, the parameters of the pointer network will be updated through the interaction with the environment, by refining the parameters of the pointer network when reasoning on a single virtual network request.

\begin{figure}[!htbp]
\includegraphics[width=1.0\columnwidth]{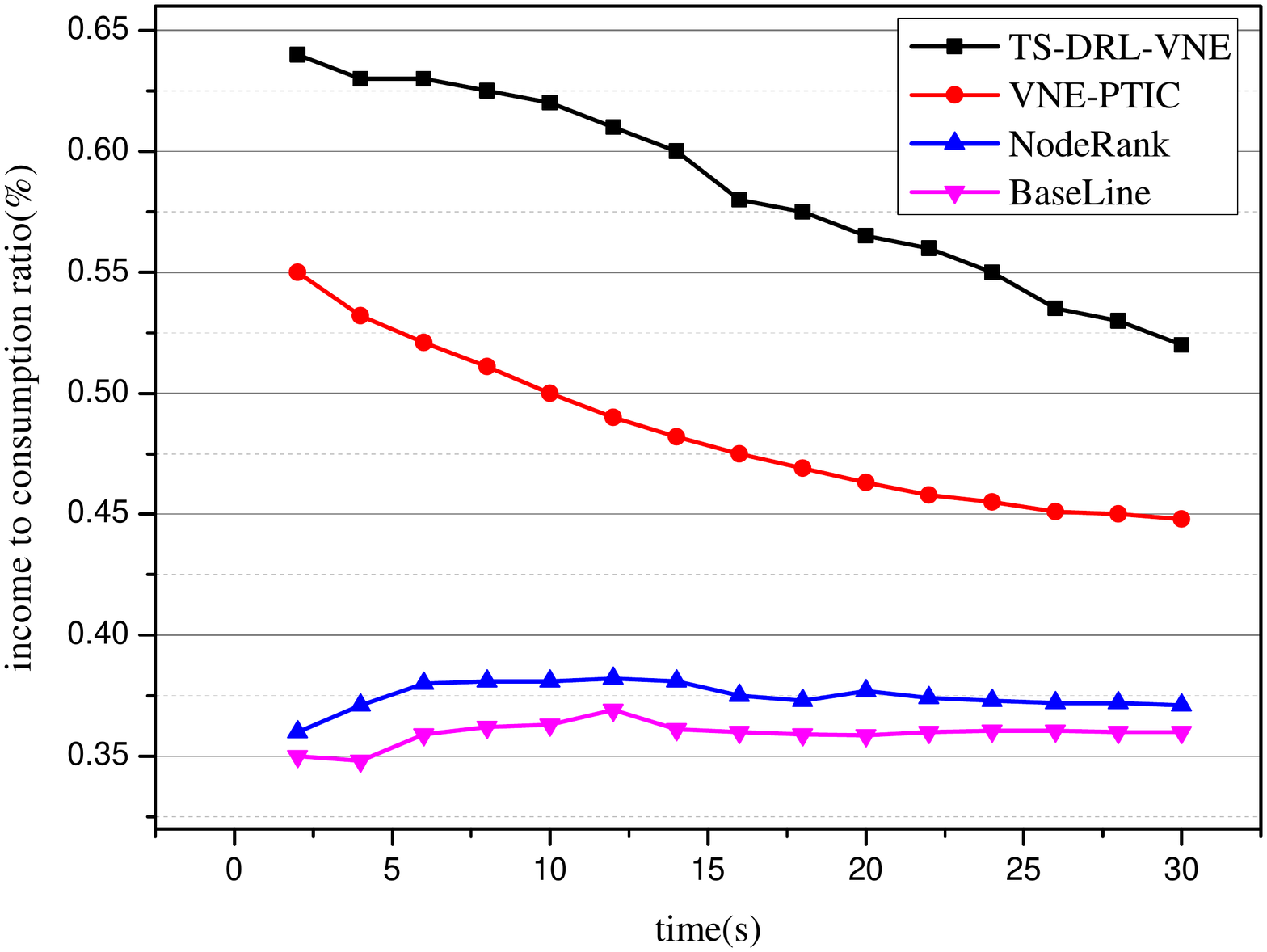}
\caption{TS-DRL-VNE income to consumption ratio.}
\label{fig_TS-DRL-VNE income to consumption ratio}
\end{figure}

Finally, we compare the TS-DRL-VNE algorithm with the VNE-PTIC algorithm proposed in reference \cite{5}, NodeRank algorithm proposed in reference \cite{B14} and  BaseLine algorithm proposed in reference \cite{B15} in terms of the income to consumption ratio. The result is shown in Figure. \ref{fig_TS-DRL-VNE income to consumption ratio}. We use hops as reward signals. Agents tend to look for solutions that use fewer hops (that is, less bandwidth resources). Therefore, the TS-DRL-VNE algorithm we proposed uses less underlying resources to accept more virtual network requests, thus obtaining higher revenue consumption ratio.

In the TS-DRL-VNE algorithm, link hops are mainly used as reward signals. The number of hops is negatively related to the size of the reward, so the agent prefers solutions with fewer hops to obtain larger rewards. In addition, learning agent can get better mapping strategy by adjusting the parameters of pointer network and training in a more reasonable network environment. Therefore, the TS-DRL-VNE algorithm uses less underlying resources to accept more virtual network requests and obtains a higher revenue to consumption ratio.

The practical significance of using this algorithm to solve the network resource allocation problem is to make the VNE algorithm more accurate and stable through the two-stage embedding of nodes and links. For example, in the application of unmanned aerial vehicle (UAV) in the military field, it is required that UAV can strike the target precisely. At the same time, with the change of target position, UAV can make decision quickly. These are all the traditional heuristic VNE algorithm cannot do. Therefore, the application of DRL algorithm to solve the problem of VNE can greatly improve the performance of the algorithm.

The proposed TS-DRL-VNE algorithm establishes a pointer network to determine the node embedding strategy. It is realized by refining the parameters of the pointer network when reasoning a single virtual network request. The performance evaluation results show that the proposed algorithm is better than the existing heuristic algorithm in the aspect of income to consumption ratio. In the future work, different reward signals can also be used to optimize agent performance to prove the performance of DRL-based VNE algorithm.

\section{DRL-VNE Algorithm Based on Full Attribute Matrix}

Most of the current VNE algorithms model virtual network and substrate network in the form of matrix. Since virtual network and substrate network contain many node and link information, attribute matrix or adjacency matrix alone is not enough to represent the whole network. Therefore, it is necessary to consider the attribute matrix and the adjacency matrix together to obtain a full attribute matrix representing the complete network. In this part, we use the method of fusion attribute matrix and adjacency matrix to change these two sparse matrices into a dense matrix, that is, the full attribute matrix of the complete network. The main advantages of this matrix processing VNE are as follows: (1) Because only similarity matrix between information is needed, the full attribute matrix is very effective for processing dimension reduction of sparse data. (2) Since the node information and link information of substrate network are fused, features of substrate network can be mined more effectively when processing dense data.

In recent years, some scholars have applied DRL to solve VNE problems. Reference \cite{6} proposed the Monte Carlo search tree algorithm. The algorithm considers the decision process of node mapping in VNE as Markov decision-making process. When the virtual request arrives, a mapping strategy is made using the Monte Carlo search tree. When the node mapping decision process is completed, the shortest path algorithm or the multi-commodity flow algorithm is used to map the link requests. Rashid et al \cite{7} also proposed a virtual network resource dynamic and opportunity allocation algorithm based on distributed Q-learning. These ML-based VNE algorithms are more efficient than heuristic algorithms. Reference \cite{8} used random search of ant colony algorithm to obtain the global optimal solution to solve the virtual network mapping problem. The ant colony algorithm first divides virtual network requests into a set of subproblems represented by solution components that are sorted and then sequentially assigned to form a sequence of mapping transformations. To do this, each ant needs to build a solution iteratively. Ant colony optimization algorithms can be VNE algorithms that do not need to assume sufficient physical resources and do not need to limit the location of physical nodes.

The above VNE algorithm based on DRL tries to improve the effect of VNE from the model direction and achieves good results. However, no breakthrough has been made in substrate network representation and training methods. They only limited to some features of virtual network and substrate network, and then trained agents in this model environment. Therefore, these algorithms do not accord with the actual VNE problem. By fusing the attribute matrix and the adjacency matrix, the substrate network attribute matrix and the adjacency matrix are fused to produce a robust full attribute matrix. Based on this matrix, FAM-DRL-VNE algorithm is proposed.

\begin{figure}[!htbp]
\includegraphics[width=1.0\columnwidth]{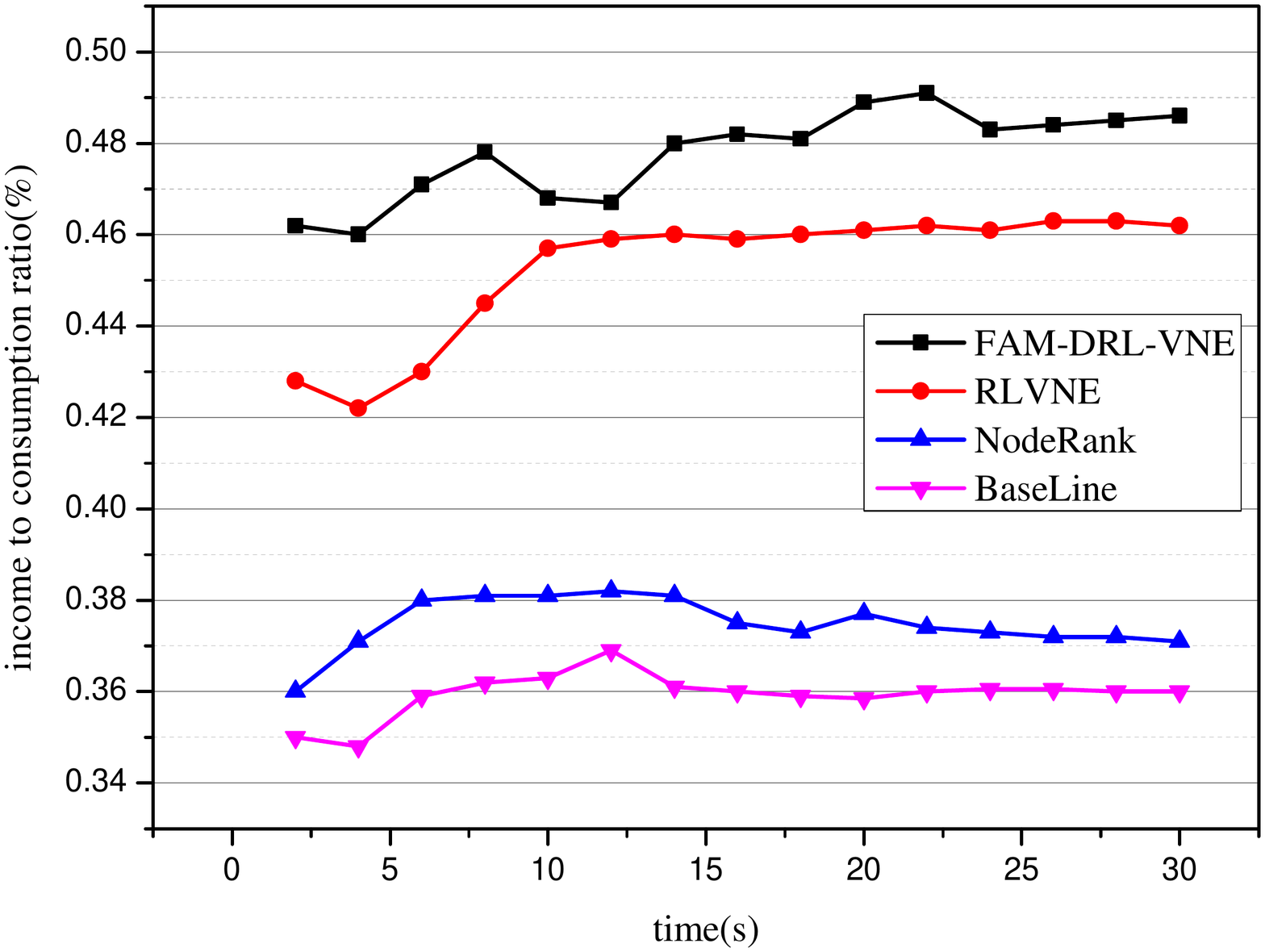}
\caption{FAM-DRL-VNE income to consumption ratio.}
\label{fig_FAM-DRL-VNE income to consumption ratio}
\end{figure}

Finally, the FAM-DRL-VNE algorithm was compared with the ML based algorithm RLVNE \cite{9}, NodeRank algorithm proposed in reference \cite{B14} and  BaseLine algorithm proposed in reference \cite{B15} in terms of the long-term income to consumption ratio. The result is shown in Figure. \ref{fig_FAM-DRL-VNE income to consumption ratio}. In the beginning, there is no significant decline in the long term income to consumption ratio, because it is independent of the amount of physical resources. Then the benefit-cost ratio leveled off, but the convergence trend and convergence value of FAM-DRL-VNE algorithm were better than that of RLVNE algorithm. From this experiment, we can draw the conclusion that the reinforcement learning agent learns the relation of substrate network nodes in FAM-DRL-VNE algorithm during the training phase, then the model can be generalized during the test phase. Finally, a good evaluation index is obtained on the test set.

The ratio of income to consumption didn't show a downward trend at the beginning, because this indicator has nothing to do with the amount of underlying resources. In the training phase, the reinforcement learning agent of the FAM-DRL-VNE algorithm is trained in an environment that is closer to the real network, and a more detailed network node relationship is learned. In the test phase, the learning agent of FAM-DRL-VNE algorithm can generalize the algorithm model matrix and fully mine the node information and link information of the substrate network. In the learning process, the agent can fully consider the relevance between each node. So in the end a higher ratio of revenue to consumption was achieved.

The practical significance of using this algorithm to solve the network resource allocation problem is to make the VNE algorithm more efficient and decision-making faster by learning the characteristics of the underlying network and efficiently training agents. For example, in the application of UAV in the field of logistics and transportation, UAV is required to be able to respond to different weather changes to deliver goods safely and on time. This requires UAVs to make decisions quickly in the face of different situations. These are all the traditional heuristic VNE algorithm cannot do. Therefore, the application of DRL algorithm to solve the problem of VNE can greatly improve the performance of the algorithm.

In the future work, the mapping relationship between virtual network and substrate network should be fully considered. More nodes and link attributes should be mined to improve the algorithm. In addition, substrate network will change after each virtual network request mapping is successful, so the substrate network feature changes dynamically at high frequency. Therefore, it is necessary to find a new and updated method to represent the feature change of substrate network.

\section{DRL-VNE Algorithm Based on Matrix Perturbation Theory}

When a virtual request is mapped to the substrate network, the parameters of the substrate network change. Therefore, in the process of VNE, the attribute matrix is a matrix with high frequency dynamic changes. If the node and link information of virtual network and substrate network is obtained through full attribute matrix for each substrate network between each virtual network request, this will cause a large time complexity. In order to solve the problem that FAM-DRL-VNE algorithm could not iterate quickly, we proposed the MPT-DRL-VNE algorithm in this part.

Perturbation theory starts with the exact solutions of related simpler problems and uses special mathematical methods to give approximate solutions to many problems that do not require exact solutions. Perturbation theory usually applies to problems with the following properties: By adding a perturbation term to the mathematical representation of the simpler part, the approximate solution of the whole problem can be calculated.

There are many examples of using DRL method to solve VNE problem. Reference \cite{10} introduced the neural network algorithm into the VNE problem. This algorithm proposes an autonomous system based on artificial neural network to improve the efficiency of VNE and realize adaptive allocation of virtual network resources. The advantage of this algorithm is that there is no need to limit the spatial dimension of input and output. Its input is the substrate network resource state, and its output is its allocation decision. In reference \cite{11}, Q-learning algorithm in ML was used to learn the optimal mapping mechanism by using the reward mechanism. Instead of allocating a fixed number of resources to the virtual request nodes throughout the virtual request cycle, the algorithm dynamically allocates substrate network resources to virtual nodes and links according to perceived needs. The contribution of this algorithm is twofold. The first is the distributed learning algorithm, which dynamically allocates resources for virtual nodes and links. Secondly, the convergence speed of the virtual network mapping model is improved by the biased learning strategy. Reference \cite{9} proposed the use of the Policy Gradient algorithm in the DRL algorithm to enable the reinforcement learning agent to gradually learn the optimal mapping mechanism. In the training phase of this algorithm, the physical node with the greatest probability is not selected by greedy algorithm, but by random selection, which means that the output may be biased and there is a better solution. This model therefore explores how to strike a balance between exploring better solutions and developing existing models.

\begin{figure}[!htbp]
\includegraphics[width=1.0\columnwidth]{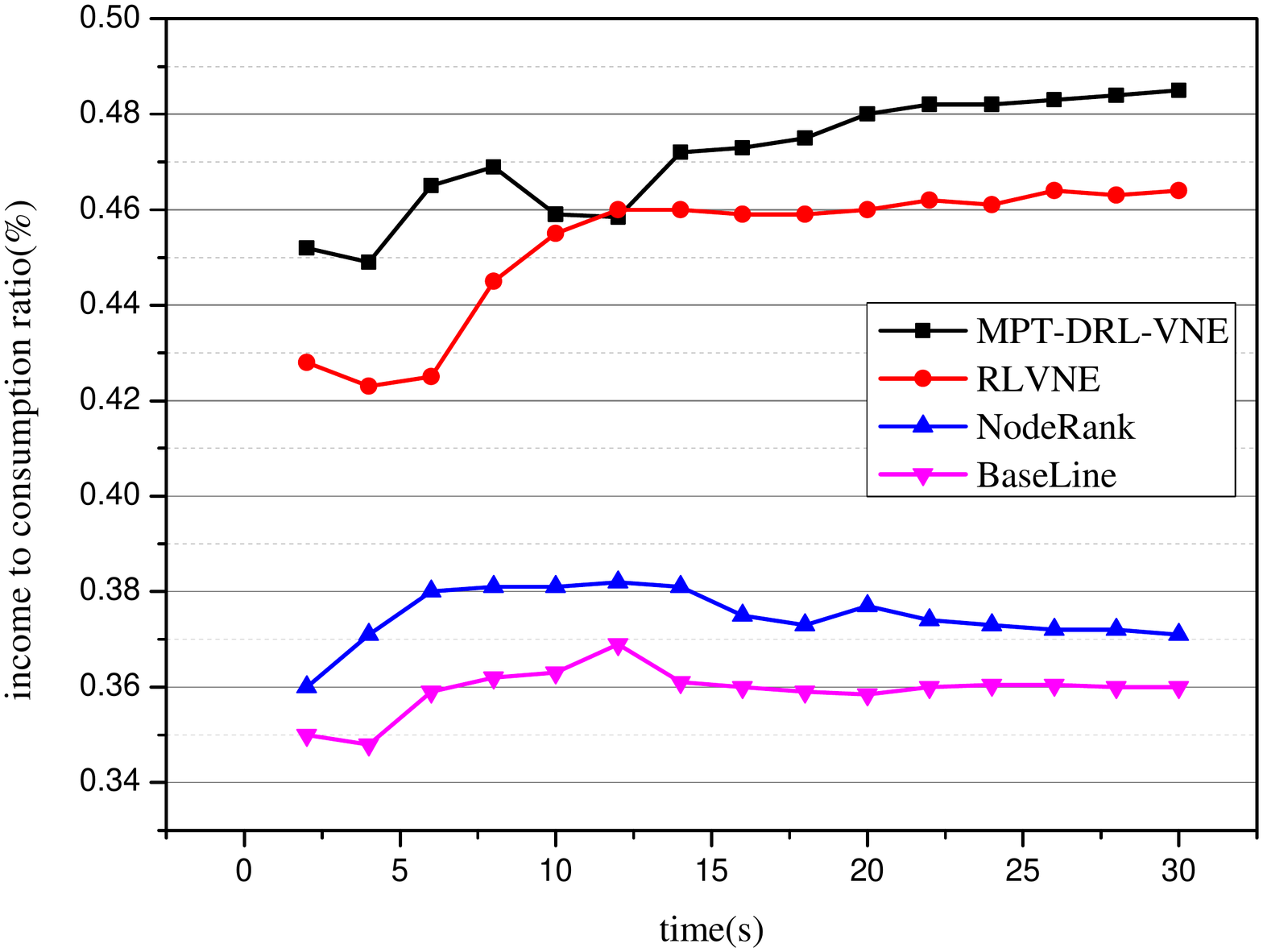}
\caption{MPT-DRL-VNE income to consumption ratio.}
\label{fig_MPT-DRL-VNE income to consumption ratio}
\end{figure}

The above DRL-based VNE algorithm does not consider the dynamic change of resources on substrate network between two VN requests, so we propose the MPT-DRL-VNE algorithm. After the full attribute matrix is obtained, for each virtual network request, the initialized full attribute matrix is dynamically updated using flow processing. This method greatly reduces the time complexity and makes it possible to obtain the consensus matrix to represent the substrate network topology through spectral analysis. Finally, we compared the algorithm with RLVNE algorithm, NodeRank algorithm proposed in reference \cite{B14} and  BaseLine algorithm proposed in reference \cite{B15} in terms of long-term income to consumption ratio, as shown in Figure. \ref{fig_MPT-DRL-VNE income to consumption ratio}. In the beginning, there is no significant decline in the long term return to consumption ratio, because it is independent of the amount of physical resources. However, the convergence trend and convergence value of MPT-DRL-VNE algorithm are better than that of RLVNE algorithm.

The ratio of revenue to consumption didn't show a downward trend at the beginning, because this indicator has nothing to do with the amount of underlying resources. In the training phase, the reinforcement learning agent of the MPT-DRL-VNE algorithm is trained in an environment that is closer to the real network, and a more detailed network node relationship is learned. In addition, in the MPT-DRL-VNE algorithm, the underlying resources between virtual network requests are constantly changing. The parameters are adjusted reasonably by spectral analysis method, so that each virtual network request can be mapped in a more realistic network environment when it arrives. So in the end a higher ratio of revenue to consumption was achieved.

The practical significance of using this algorithm to solve the problems faced in the field of network resource allocation is that the agent can adjust the strategy to better learn the actual situation of the next network request by understanding the resource changes between each virtual request. For example, in the application of sensors in the field of environmental monitoring, it is required that the sensors can respond to changes in different seasonal conditions and pass back information such as temperature and humidity at any time. This requires smart sensors to make quick decisions when faced with different situations. These are the traditional heuristic VNE algorithm cannot do. Therefore, the application of DRL algorithm to solve the VNE problem can greatly improve the performance of the algorithm.

The index of evaluating VNE algorithm is not limited to the ratio of income to consumption. Such as mapping cost, virtual network request acceptance rate and so on are also important criteria to judge VNE algorithm. Therefore, the algorithm can be optimized in these aspects in the future.

\section{Conclusion}

In response to the problem that the rapid development of modern technologies in the network field puts forward new requirements on the resource allocation capabilities of the underlying network architecture, this paper uses a new network architecture -- NV technology to propose several different solutions, focusing on solving the problem of VNE in NV.

At present, the main method to solve VNE problem is heuristic algorithm. It is worth noting that existing researchers have combined DRL algorithm with VNE problem and achieved good results. What cannot be ignored is that these methods still have problems such as easy convergence to local optimal solution, ignoring the importance of substrate network representation and training mode, and ignoring the change of underlying resources in the stage of virtual network request connection. To solve the above problems, this paper proposes three algorithms, TS-DRL-VNE, FAM-DRL-VNE and MPT-DRL-VNE. The experimental results show that these algorithms can achieve better income to consumption ratio than existing algorithms.

In the future, more efficient and reliable network models and algorithms can be explored to train agents. Considering the actual situation of the network, the performance of the VNE algorithm is further improved.

\section*{Acknowledgments}
This work is partially supported by the Major Scientific and Technological Projects of CNPC (Grant No. ZD2019-183-006), and partially supported by ``the Fundamental Research Funds for the Central Universities" of China University of Petroleum (East China) (Grant No. 20CX05017A, 18CX02139A). The authors also gratefully acknowledge the helpful comments and suggestions of the reviewers, which have improved the presentation.

% Can use something like this to put references on a page
% by themselves when using endfloat and the captionsoff option.
\ifCLASSOPTIONcaptionsoff
  \newpage
\fi

%\bibliography{references}

\end{document}